\documentclass[12pt]{JHEP3}
\input epsf.tex
\epsfclipon

\usepackage{epsfig}

\usepackage{bbm,bm,amsmath,amssymb}

\def\be{\begin{equation}}
\def\ee{\end{equation}}
\def\baray{\begin{eqnarray}}
\def\earay{\end{eqnarray}}

\newcommand{\roughly}[1]{\mathrel{\raise.3ex\hbox{$#1$\kern-0.85em
\lower1ex\hbox{$\sim$}}}}
\def\lsim{\roughly<}

\def\2pi{\left(2\pi\right)}

\def\beq{\begin{equation}}
\def\eeq{\end{equation}}
\def\beqa{\begin{eqnarray}}
\def\eeqa{\end{eqnarray}}

\def\Hod{H^{{\phantom{|}^{\phantom{|}}}^{\!\!\!\!\!\!\!\!\!\circ}}}
\def\cHod{{\cal H}^{{\phantom{|}^{\phantom{|}}}^{\!\!\!\!\!\!\!\!\!\circ}}}
\def\pod{\phi^{{\phantom{|}^{\phantom{|}}}^{\!\!\!\!\!\!\!\circ}}}
\def\cHodd{{\cal H}^{{\phantom{|}^{\phantom{|}}}^{\!\!\!\!\!\!\!\!\!\!\!\circ\;\!\circ}}}
\def\podd{\phi^{{\phantom{|}^{\phantom{|}}}^{\!\!\!\!\!\!\!\!\!\circ\;\!\circ}}}
\def\cH{{\cal H}}
\def\cP{{\cal P}}

\title{Inflationary Potential Reconstruction for a WMAP Running Power Spectrum}

\author{James M.\ Cline, Loison Hoi\\
Physics Department, McGill University \\
3600 University Street, Montr{\'e}al \\
        Qu{\'e}bec, Canada, H3A 2T8}

\date{\today}

\abstract{The {first year} WMAP measurement of the CMB temperature
anisotropy is  intriguingly consistent with a larger running of the
inflationary scalar spectral index than would be expected for
single-field inflation.  We revisit the issue of a large running
spectral index, first by reexamining the evidence from the data, and
then by  reconstructing the inflationary potential, using an improved
method based upon the Hamilton-Jacobi formulation.  We note that a
spectrum which runs only over  1.5 decades  of $k$ space provides as
good a fit to the CMB data as one which runs at all $k$, that
significant evidence for running comes from multipoles {$l$ near 40},
and that  large running gives a better fit than a flat spectrum 
primarily  if the tensor-to-scalar ratio $r$ is large, $r\sim 0.5$,
and the field values are at the Planck scale.  This allows one to
break the large degeneracy of potentials which would  be consistent
with the scalar power alone. Large running, should it be confirmed,
is thus linked to a high scale of inflation and the  possibility of
seeing effects of tensor modes in the CMB {and Planck-scale
physics}.  Nevertheless, we show that the reconstructed inflaton
potential is well-described by a  renormalizable potential whose
quantum corrections are under  control despite the large field
values. }

\maketitle

\begin{document}

\section{Introduction}
\label{intro}

In 2003 the Wilkinson Microwave Anisotropy Probe (WMAP)
released the results
of its first year of high-precision measurements of the temperature
anistropy of the Cosmic Microwave Background (CMB) \cite{Bennett}.  While
several  anomalies of marginal statistical significance have been
noted,  one of the most visually striking is the trend of the
spectral index of the scalar power $n_s$, 
to run toward smaller values \cite{Peiris},
shown in figure \ref{fig1a}.  
There it was determined that ${d n_s/ d\ln k}\sim-0.1$ albeit
with large error bars.
Like the other anomalies,
this one also remains to be proven statistically significant, since a
flat spectrum is also consistent with the data. Nevertheless, since
the present data have opened the door to the possibility of large
running, it is interesting to explore the theoretical consequences
while we continue to wait for the next WMAP data release.

\DOUBLEFIGURE{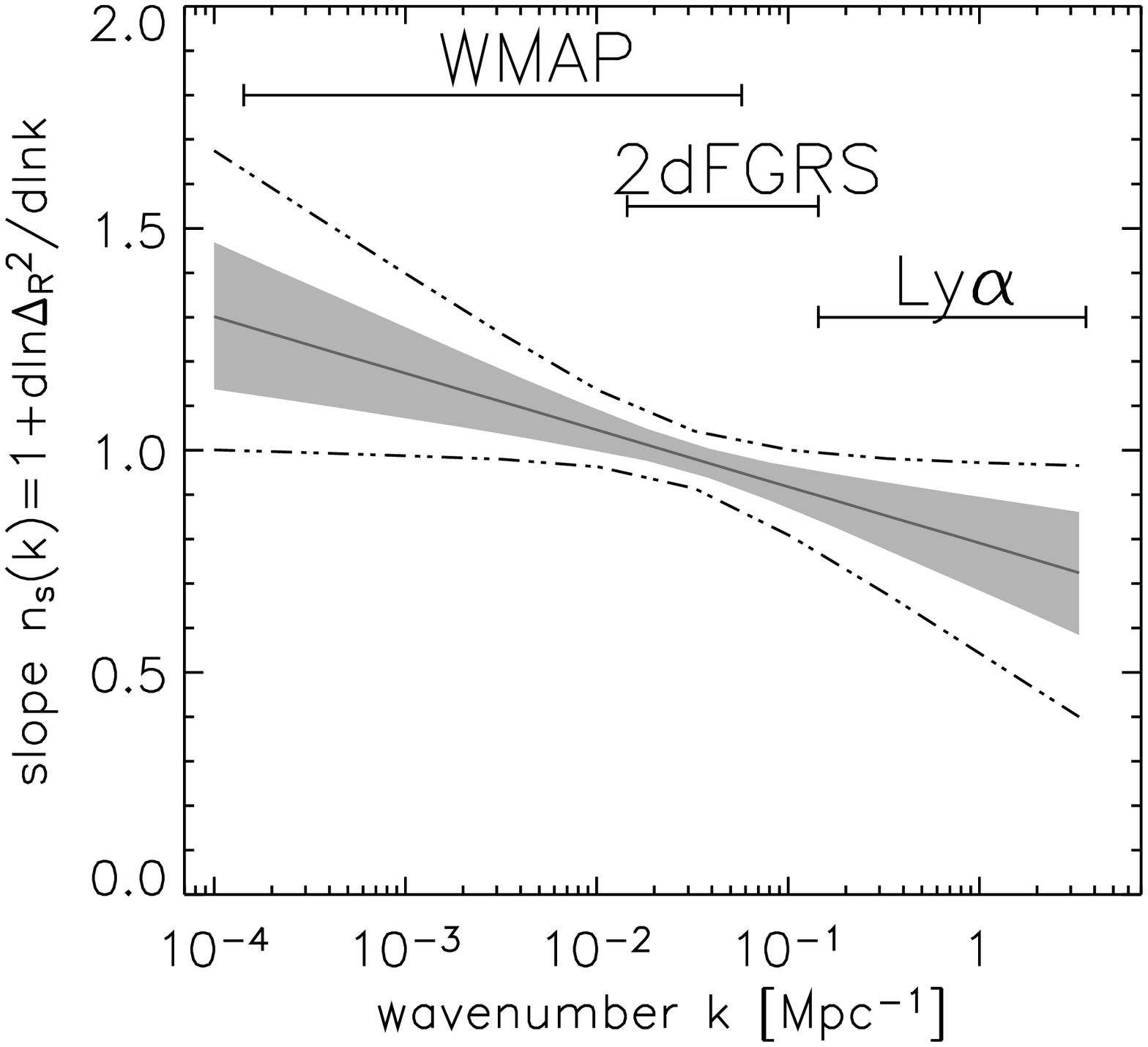,width=\hsize}
{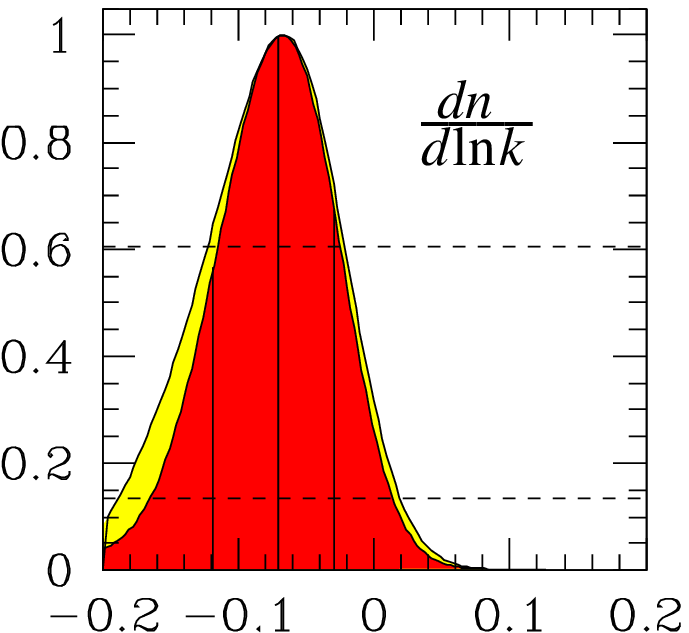,width=\hsize}{Experimental limit on the
scale-dependent spectral index, reproduced from figure 2 of
\cite{Peiris}.  Shaded areas and dashed lines indicate 68\% and
95\% confidence regions, respectively. \label{fig1a}}{Probability
distribution function for 
running from combined WMAP and SDSS data, from \cite{SDSS-teg}.
\label{fig1b}}


It is somewhat surprising that the issue of running has not attracted
as much attention as the low quadrupole since it is arguably as 
significant.  Nevertheless there have been a number of papers 
which have treated running, starting even before WMAP
\cite{KT}.  Refs.\ \cite{HH1, HH2} were the first to examine
the likelihood of a running spectrum using modern CMB and Lyman
$\alpha$ data.
Subsequently to WMAP, ref.\ \cite{BLWE} found that the marginal evidence
for running disappeared if one omitted the first three multipoles,
a conclusion which we shall challenge below.
Including data from the Sloan
Digital Sky Survey galaxy survey \cite{SDSS-teg} slightly narrows
the errors on ${d n_s/ d\ln k}$, but not the central value,
as shown in figures \ref{fig1a}-\ref{fig1b}.

From the theoretical perspective, the $-0.1$ level of running 
 is uncomfortably large, since ${d n_s/ d\ln k}$ is a
second order effect in the slow-roll expansion; generically one
expects ${d n_s/ d\ln k}\lsim 1/N_e^2$ with $N_e\sim 60$
e-foldings of inflation since horizon crossing.  Ref.\ \cite{CST}
considered which general classes of models could give large running,
and methods of constructing potentials with large running.
Refs. {\cite{recon1}-\cite{Cortes}} showed that there is a new
consistency condition between the scalar and tensor power spectra in
the presence of running.  Ref.\ 
\cite{easther} analyzed  generically large running in conjunction
with multifield inflation, and sudden changes of direction in field
space.  Supergravity models with two-stages of inflation have
been shown to provide large running \cite{yam-yok}.  Classes of
well-motivated particle physics models consistent with large
running were presented in \cite{BCE}; inflation in
noncommutative spacetime has been found to give significant
running \cite{BH, HL}, as well as inflation with a cutoff
on transPlanckian fluctuations \cite{Ash}.  
  In \cite{CS} it was noted
that although brane-antibrane inflation models predict smaller
levels of running, $dn_s/d\ln k\sim -0.01$, nevertheless the
correlation of $dn_s/d\ln k$  with $n_s$ can provide an important
test of the model if future experiments are able to reliably 
measure running at this low level.

One possible reason for a decline of interest in running of $n_s$
is that recent Lyman-$\alpha$ forest data have moved the central
value much closer to zero, and narrowed the confidence intervals.
Ref.\ \cite{sdss-sel} finds that $dn_s/d\ln k = -0.003\pm 0.010$
(see also \cite{popa}).
This determination would seem to put the question of large running
to rest.  However, given the reasonable question as to whether systematic
errors are completely under control in the  
Lyman-$\alpha$ studies, we suspend
judgment until there is further confirmation from WMAP or
other CMB experiments.  Moreover,
Lyman-$\alpha$ data are sensitive to the smallest scales.  It is
perfectly possible that the power spectrum runs significantly at
larger scales, but flattens out at smaller scales.  Indeed, this 
is almost a necessity since it is impossible to sustain inflation over
60 e-foldings if the running remains large throughout, as we will
discuss.

There is a broader issue connected with increasingly precise
experimental determinations of the inflationary power spectrum,
namely, what is the shape of the inflaton potential which can
generate a given spectrum {\cite{Hodges, recon}}?  One of the points
of this paper is that an expansion of the potential around a point
using slow-roll parameters, and integrating the slow-roll
approximation of the inflaton equations of motion, are inadequate for
accurately reproducing the desired spectral shape, if it runs at the
level which we are considering, or otherwise departs from the
slow-roll approximation.  In this work we present a simpler technique
for reconstructing the potential, based on the Hamilton-Jacobi
formulation of the inflaton equation of motion \cite{Muslimov,
SalopekandBond}.  

We begin in section \ref{rps} with a description of the running
spectrum and its likelihood, with our own analysis of CMB and 
large scale structure (LSS) data sets.  We argue that the possibility
of large running, though unproven, is still plausible, if applied to
a restricted range of the power spectrum.
In section \ref{recon} we
describe our new method of reconstructing the potential, based on the
Hamilton-Jacobi method.  A one-parameter family of potentials
corresponding to the running scalar power spectrum is constructed.  
In section \ref{src} we show how higher {order} corrections in the slow-roll expansion 
for the power spectrum can be efficiently incorporated.
In section \ref{sic} we note that a given potential yields the desired
power spectrum only for a specific initial velocity for the inflaton,
which does not generally correspond to the slow-roll attractor value. 
In section \ref{tc} we break the degeneracy between potentials found in the
previous section by assuming a large tensor-to-scalar ratio, as is favored
by the best fit parameters found in section \ref{rps}.  We discuss 
several aspects of this potential, including the duration of inflation
which can be supported during the period of large running, and the 
extent to which the potential can be approximated by renormalizable
terms in a quantum field theory.
We give conclusions in section \ref{conc}.

\section{Running Power Spectrum}
\label{rps}
The power spectrum with running spectral index is parametrized as
\beq
\label{power}
	P(k) = A\, \exp\left[(n_s-1) \ln (k/k_0) + \frac12 {dn_s\over d\ln k} 
	\ln^2(k/k_0) \right] 
\eeq
The running spectrum which best fits the {first year} WMAP data was given in
\cite{Peiris} with
\beq
	n_s = 1.27,\quad {dn_s\over d\ln k} = -0.1
\eeq
at a pivot point of $k_0 = 0.002$ Mpc$^{-1}$, with a likelihood given
by $-2\ln{\cal L}= \chi^2 = 1426$ for 1340 degrees of freedom (number
of data points minus number of parameters).   The other cosmological
parameters are listed in table \ref{tab2}, where we show the best-fit
running and nonrunning
models which we obtain using the CosmoMC Markov Chain Monte Carlo
program {\cite{cosmomc}} (see http://cosmologist.info/cosmomc).  The $\chi^2=1426$ value
for the running spectrum includes a penalty of $\Delta\chi^2 = {1.4}$
from the high value of the Hubble parameter, $h\cong {0.814}$, relative
to the Hubble Space Telescope value $h\cong 0.72$ {\cite{HST}}.\footnote{
Nevertheless our value
is somewhat smaller than the value obtained by WMAP for the
running model \cite{Peiris}, $\chi^2=1428$.}
All models shown
are flat: $\Omega_{\Lambda} =  1 - \Omega_m$.  We note that a
sizeable tensor-to-scalar ratio $r$  (as well as a large optical
depth $\tau$) is required to get the best fit with large running.  
If we set $r=0$ while keeping $n_s$ and  ${dn_s / d\ln k}$ fixed (but
allowing other cosmological parameters to vary) the likelihood
decreases, $\chi^2 = 1450$. There is a degeneracy between running and
tensors; if we set  $r=0$ and allow all other parameters to vary,
including $n_s$ and $dn_s / d\ln k$,  the best fit has $\chi^2 =
1429$, but smaller running, $dn_s / d\ln k \sim -0.009$.  Although
the statistics are not yet significant, there is the hint that large
running and a large tensor ratio fit the data better.



\TABLE{\centering
\begin{tabular}{|c|c|c|c|c|c|c|c|c|}
\hline
$n_s$ & ${dn_s\over d\ln k}$ $\phantom{1_|^|\!\!\!\!\!\!}$ & $r$ & $A$ & $\Omega_b h^2 $ & $\Omega_m h^2$ & $h$ & $\tau$ &
$\chi^2/\nu$\\
\hline
1.27 & $-0.1$ &0.57 & $2.1\times 10^{-9}$ & 0.024  & 0.122 & 0.81 &
0.29  & 1426/1340\\
0.99 & $-$ & $-$& $2.6\times 10^{-9}$ & 0.024 & 0.140 & 0.72 & 0.16 & 1429/1342\\
\hline
\end{tabular} \label{tab2}\caption{Our best fits for the 
running and nonrunning
models using CosmoMC and the {first year} WMAP data.}
}

We have repeated the comparison between running and nonrunning models using
data from other CMB experiments (ACBAR \cite{ACBAR}, VSA \cite{VSA}, CBI \cite{CBI}), and galaxy surveys
(2dF \cite{2dF}, SDSS \cite{SDSS-teg}).  (In all the data sets we also include the 
Hubble Space Telescope determination of the Hubble parameter, which
contributes $156.25 \times (h - 0.72)^2$
to the $\chi^2$, leading to  $\Delta\chi^2 \simeq {1.4}$ between the running
and nonrunning models using the WMAP data.)
The best fit parameters for different combinations of these
data are shown in table \ref{tab3}.  Inclusion of other CMB data 
strengthen the evidence for running, while including 
large scale structure (LSS)
data weaken the evidence.  This is more evident with the inclusion of Sloan
Digital Sky Survey (SDSS) data than with the 2dF data.\footnote{We have observed
that there is a tendency for CosmoMC to get
stuck in local minima of $\chi^2$ when searching for the best fit
parameters, especially when SDSS is included, so it is possible that we have not found
the global minimum $\chi^2$.}
This is consistent with the
evidence for running coming principally from the low-$k$ part of the power
spectrum, to which the CMB is more sensitive.  If the effect turns out to be real,
then the best fit to the data including large scale structure will require a
spectrum which has some different behavior
from eq.\ (\ref{power}) at large $k$, as we will discuss below.

\TABLE{\centering
\begin{tabular}{|c|c|c|c|c|c|c|c|c|c|}
\hline
data & $n_s$ & ${dn_s\over d\ln k}$ $\phantom{1_|^|\!\!\!\!\!\!}$ & $r$ & $10^9\,A $ & $\Omega_b h^2 $ & $\Omega_m h^2$ & $h$ & $\tau$ &
$\chi^2/\nu$\\
\hline
WMAP & 1.27 & $-0.10$ &0.57 & $2.1$ & 0.024  & 0.122 & 0.81 & 0.29  & 1426/1340\\
& 0.99 & $-$ & $-$& $2.6$ & 0.024 & 0.140 & 0.72 & 0.16 & 1429/1342\\
\hline
CMB & 1.27 & $-0.095$ &0.57 & $2.1$ & 0.025  & 0.127 & 0.80 & 0.29  & 1447/1363\\
& 0.97 & $-$ & $-$& $2.5$ & 0.023 & 0.131 & 0.74 & 0.14 & 1453/1365 \\
\hline
2dF & 1.09 & $-0.047$ &0.24 & $2.3$ & 0.023  & 0.147 & 0.69 & 0.13  & 1462/1372\\
& 0.96 & $-$ & $-$& $2.5$ & 0.023 & 0.142 & 0.69 & 0.11 & 1464/1374 \\
\hline
CMB& 1.09 & $-0.051$ &0.23 & $2.3$ & 0.023  & 0.144 & 0.69 & 0.13  & 1485/1395\\
+2dF& 0.96 & $-$ & $-$& $2.5$ & 0.023 & 0.140 & 0.71 & 0.11 & 1488/1397\\
\hline
SDSS & 1.09 & $-0.039$ &0.24 & $2.3$ & 0.023  & 0.153 & 0.68 & 0.14  & 1456/1359\\
& 0.99 & $-$ & $-$& $2.6$ & 0.024 & 0.160 & 0.67 & 0.13 & 1454/1361\\
\hline
mpk & 1.09 & $-0.036$ &0.24 & $2.3$ & 0.024  & 0.155 & 0.68 & 0.14  & 1490/1391\\
& 0.99 & $-$ & $-$& $2.6$ & 0.024 & 0.152 & 0.69 & 0.15 & 1490/1393\\
\hline
CMB& 1.09 & $-0.045$ &0.23 & $2.3$ & 0.023 & 0.148 & 0.68 & 0.14  & 1517/1414\\
+mpk& 0.96 & $-$ & $-$& $2.5$ & 0.023 & 0.149 & 0.68 & 0.09 & 1517/1416\\
\hline
\end{tabular}
\label{tab3}
\caption{Best fit parameters for both running and nonrunning
spectral index models from CosmoMC. The first row is the result from
the {first year} WMAP data alone; others are from the combinations of the WMAP
data and the data sets specified at the first column, where
CMB=ACBAR+CBI+VSA, mpk=2dF+SDSS.
We used the July 2005 version of CosmoMC, and its
default data sets. They are: the first year WMAP data, $2 \le l_{\rm
TT} \le 900$, $2 \le l_{\rm TE} \le 450$ (1348 data points); July
2002 ACBAR data \cite{ACBAR}, $900 \le l \le 1950$ (7 data points);
2000 and 2001 CBI data \cite{CBI}, $700 \le l \le 1760$ (8 data
points); Feb 2004 VSA data \cite{VSA}, $640 \le l \le 1700$ (8 data
points); June 2002 2dF data \cite{2dF}, 32 data points;
Oct 2003 SDSS data \cite{SDSS-teg}, $0.016 \le k(h/{\rm Mpc}) \le 0.205$
(19 data points).}
}

\DOUBLEFIGURE[h]{partial.eps,width=2.8in}{multipole2.eps,width=2.8in}{Log of best fit 
running spectrum
(solid, black) and the partial running spectrum (dashed, red) versus $\ln k$. 
The nonrunning spectrum is also shown (dot-dashed, green). 
\label{fig2}}{Difference of $\chi^2$ between the 
{nonrunning} and {running} models, versus {the minimum multipole included $l_{\rm min}$}; 
see eq.\ (2.3). The separate contributions from temperature (TT) and
polarization (TE) as well as the total are shown. 
\label{multipole}} 

The relevant part of the spectrum is shown in figure \ref{fig2}; the
range of $k$ values here is that which is actually called by the CAMB
program (see  http://camb.info/) in computing the multipole moments
of the temperature anisotropy and polarization of the WMAP data. The
scales $k=10^{-4}$ Mpc$^{-1}$ and $k=10^{-1}$ Mpc$^{-1}$ indicated on
figure \ref{fig2} correspond roughly to the
quadrupole and the $l=2000$ mode of the multipoles of the acoustic
peaks.  It is interesting to note that the {first year} WMAP data  are
equally consistent with both curves shown in figure \ref{fig2}; the
solid one having a constant value of $dn_s / d\ln k$, and the dashed
one having  ${dn_s / d\ln k}=0$ outside of the range ${-7.2}  < \ln
(k\cdot{\rm Mpc}) < {-3.8}$.  Both of these spectra have the same
likelihood, even without varying the cosmological parameters given in
table \ref{tab2}.  We have verified this using our modified versions
of CAMB  and CosmoMC which allow for an arbitrary user-specified
power spectrum.  We will make further use this truncated version
of the running spectrum, which is flat at large and small $k$,
referring to it as the {\it partial running} model.

It is at first surprising that the partial running model should
provide as good a fit to the data as the constant running model,
since \cite{BLWE} showed that the evidence for running comes  mainly
from the lowest {three} multipoles.  In contrast the running part of our
partial running spectrum affects much higher multipoles.  To 
better understand this, we have computed the difference in
$\chi^2$ between the {nonrunning} and {running} models including  
contributions {from $l_{\rm min}$ to $l_{\rm max}$ (900 for TT and 450 for TE):
\beq
	\label{lmin}
	\Delta\chi^2 (l_{\rm min}) = \sum_{l=l_{\rm min}}^{l_{\rm max}} \delta\chi^2(\rm nonrunning)-\delta\chi^2(\rm running)
\eeq
where $\delta\chi^2(l)$ is the $\chi^2$ contributed from diagonal ($l$) and off-diagonal $(l^\prime>l)$ terms.}
This is shown in figure \ref{multipole}.\footnote{Since the original
submission of this paper, we have reanalyzed the data using the
WMAP three-year data release.  The shape of the total $\Delta\chi^2$
curve is unchanged, but it has come down by approximately
$\Delta\chi^2=1$ due to a similar shift in the contribution from 
TE.}  We observe there that the low multipoles ($l=2,3,4$) 
in the TT data do account for
a substantial part of the total decrease in $\chi^2$ between the nonrunning
and running models; however the TE data, ignored by the analysis of
\cite{BLWE}, have the opposite effect
{and hence the $\Delta\chi_{\rm tot}^2$ is rather insensitive
to the inclusion of the first three multipoles.}
In fact, the larger part of the
decrease in $\chi^2$ clearly comes from the region near $l=40$
 {in the
TT data}, which 
explains why our partial running spectrum starts running at relatively
high $k$ values, $k\cong 0.001$ Mpc$^{-1}$, compared to the the values
which would affect the low multipoles, $k\cong 10^{-4}$ Mpc$^{-1}$.
This perspective lends more interest to the possible confirmation
or negation of large running by future improvements in the data, since
the experimental determination of the higher multipoles is not so
limited by cosmic variance.

Although the LSS data do not favor the running models as shown in
table $\ref{tab3}$, we have also looked for the best fit models using
the partial running spectrum, where we allowed the  low and high $k$
values which define the partial running model to  vary.\footnote{
In
trying to understand why the SDSS analysis \cite{SDSS-teg} marginally
favors running, yet CosmoMC using SDSS data does not, we note that
the SDSS analysis uses a non-flat model when running is considered.
}
Table \ref{tab4}
shows the best fit values and the likelihood for the different data
sets.
Since other CMB data strengthen the evidence for running, but LSS data
weaken it, when including LSS data, the lower cutoff will be larger than, and the higher cutoff will be smaller than, those including other CMB data.
All the $\chi^2$ values for the partial running model are seen to
be lower than or equal to the running and nonrunning models.

\TABLE{\centering
\begin{tabular}{|c|c|c|c|c|c|c|c|}
\hline
&WMAP & CMB & 2dF & CMB+2dF &SDSS&mpk & CMB+mpk\\
\hline
$\ln k_{\rm low}$&$-7.2$&$-7.1$&$-6.5$&$-6.4$&$-6.5$&$-6.5$&$-6.5$\\
\hline
$\ln k_{\rm high}$& $-3.8$ &$-2.4$ & $-2.8$&$-2.4$ &$-2.8$  &$-2.7$&$-2.6$ \\
\hline
$\chi^2$ & 1426 & 1447&1462&1485&1454&1488&1514\\
\hline
\end{tabular} \label{tab4}\caption{The likelihood of the partial running model, using the data
described in table \ref{tab3}.}
}


To summarize our view of the experimental evidence: there is an
interesting hint of a running spectrum which is driven in large
part by multipoles near $l=40$ in the CMB data; the inclusion of 
large scale structure data do not reinforce this, but they also do not
contradict it.  Although the statistical significance is not great,
it can also be argued that it is not negligible.  According to the
Bayesian information criterion \cite{Liddle}, one should decide whether
the addition of new parameters is justified by considering whether
the combination
\[	BIC = \chi^2 + k\ln N \]
is decreased, where $k$ is the number of parameters and $N$ is the 
number of independent data points.  Not all multipole moments of the
CMB should be considered as independent, since it is known that the
acoustic peaks can be fit with great accuracy by a Fourier series with
20 terms.  A change of $\chi^2$ of $3$ can therefore support $3/\ln 20
\cong 1$ new parameters.  Although we introduced two new parameters,
the running and the tensor-to-scalar ratio, there is a degeneracy between these,
which makes them more like one parameter.  The effect we are
discussing is therefore on the borderline of being significant,
a situation that could change in either direction with new data.

\section{Reconstruction of Potential}
\label{recon}

\subsection{Formalism}

To find inflaton potentials corresponding to a desired power
spectrum,  the Hamilton-Jacobi formulation of inflation is useful 
{\cite{Muslimov, SalopekandBond}}. {Here we introduce a method
based on integration over $H(k)$, which is simple and accurate; see
ref.\ \cite{recon2} for more discussion and results in constant
spectral index case. Ref.\ \cite{Hodges} presented
 a similar integration
method, but used the lowest order slow-roll approximation in
each step, and  hence their method is not as accurate as
Hamilton-Jacobi formulation which we describe.} 
Initially, one takes the Hubble
parameter to depend on the inflaton field, $H=H(\phi)$.  In units
where $M_{\rm Pl}=1$, the Friedmann equation can be written
\beq
\label{HJ}
	3H^2 - 2H'^2 = V
\eeq
where $H' = dH/d\phi$.  The inflaton velocity is given by 
\beq
\label{phid}
	\dot\phi = - 2H'
\eeq

We begin by using the lowest order prediction for the power spectrum of the 
curvature (scalar field) perturbation in the slow-roll approximation:
\beq
\label{Pk0}
	P(k) = {H^4\over c_1 \dot\phi^2}
\eeq
where $c_1 = 4\pi^2$, and the $k$ dependence is implicitly defined using the 
horizon crossing condition,
\beq
\label{hc}
	k/a = H \ \longrightarrow \ \ln k = \ln a + \ln H
\eeq
Since, using (\ref{phid}), $\ln a = \int^t dt H = \int^\phi d\phi H/\dot\phi = -\frac12
\int^\phi d\phi\, H/H'$ , one can take $d/d\phi$ of eq.\
(\ref{hc}) to obtain a first order differential equation determining $k(\phi)$:
\beq
\label{dlnkdphi}
	{d\ln k\over d\phi} = {H'\over H} - {H\over 2 H'}
\eeq

Our strategy for reconstructing the potential is to rewrite the above equations
using $\ln k$ as the independent variable instead of $\phi$.  By straightforward
manipulation using the chain rule, and eliminating $\dot\phi$ from eqs.\
(\ref{phid}-\ref{Pk0}), we find that
\beqa
	\Hod \equiv {d H\over d\ln k} &=& -{H^3\over {2 c_1}P - H^2} \nonumber\\
	\pod \equiv {d\phi\over d\ln k} &=& \pm
	\left[ -{2\Hod\over H} \left( 1 - {\Hod\over H}
	\right) \right]^{1/2}
	=\pm \frac{ 2H \sqrt{c_1P}}{2 c_1P - H^2}
\eeqa
{and we choose the plus sign throughout this paper.  This choice of sign 
has no physical significance during any period when $\phi$ is changing monotonically
with time, as during inflation.}
These equations can be integrated numerically to find $H$ and $\phi$ as functions
of $\ln k$.  This implicitly defines $H(\phi)$, which can then be used in (\ref{HJ})
to find $V(\phi)$.  There is one relevant constant of integration 
$H_0$, the initial value of $H$, which determines the overall inflationary energy scale.
The initial value of $\phi$ has no physical importance.  {For example we can take
$\phi=0$ when $k$ is at the pivot point $k_0$ at which the power spectrum is normalized
in (\ref{power})}.

In the following, it is convenient to refer to rescaled variables.  We define
a rescaled power spectrum
\beq
	P = A {\cal P}
\eeq
where $A$ is the amplitude 
of the spectrum in eq.\ (\ref{power}), and thus ${\cal P}$ is the 
exponential factor ${\cal P}=\exp [ (n_s-1)\ln {k\over k_0} +
\frac12 \frac{dn_s}{d\ln k}\ln^2\frac{k}{k_0}]$.   
$A$ is determined by the COBE normalization, which gives
{$ P(k_c) = 25/4 \times (1.91 \times 10^{-5})^2 \simeq 2.3 \times 10^{-9}$
at $k_c = 7.5a_0H_0 \simeq 0.002{\rm Mpc}^{-1}$, 
consistent with the values in table \ref{tab2}.}
We further rescale 
\beq
\label{alphaeq}
	H = \alpha {\cal H}; \quad \alpha^2 = 2 c_1{A}
\eeq
The equations of motion become
\beqa
\label{scaled}
	\cHod &=& -{{\cal H}^3\over {\cal P} - {\cal H}^2} \nonumber\\
	\pod  &=& 
	 {\cH\,\sqrt{2{\cal P}}\over \cP-\cH^2}
\eeqa
Similarly, we define the rescaled potential through
\beq
	V = \alpha^2 {\cal V} = \alpha^2 \left[ 3{\cal H}^2 - 2{(\cHod/\pod)}^2\right]
	= \alpha^2 {\cal H}^2 \left( 3 - \frac{{\cal H}^2}{\cal P} \right)
\eeq

The Hamilton-Jacobi slow-roll parameters are
\beqa
	\epsilon &=& 2\left( {H^\prime \over H} \right)^2\\
	\eta &=& 2\left( {H^{\prime \prime} \over H} \right)
\eeqa
From eqs.\ (\ref{phid} - \ref{Pk0}), we can rewrite the $\epsilon$ as
\beq
	\epsilon = {{\cal H}^2 \over {\cal P}}
\eeq
Since $\epsilon<1$ is required for accelerated expansion,  
${\cal H}^2$ is always smaller than ${\cal P}$ during inflation,
so the equations of motion (\ref{scaled}) remain nonsingular.

\subsection{Reconstructed Potentials}
\label{RP}

The method described above can be used to numerically determine a family of
potentials $V(\phi)$ which give rise to a desired scalar power spectrum
$P(k)$.  There is no unique potential for a given $P(k)$ since changing the
scale of inflation can be compensated by a change of the slope of the potential.
This can be seen through the relation $P\sim V^3/V'^2$.  Later on we will
break this degeneracy by invoking information about the tensor power spectrum.

As a first step, we reconstruct the potentials only over the region of field
space traversed by the inflaton during the first $\sim 8$ e-foldings of inflation after
horizon crossing, since this is the relevant period for affecting the CMB.
We numerically integrated the equations (\ref{scaled}) subject to the initial
conditions $\cH = \cH_0$, $\phi=0$.  To reproduce the spectrum shown in figure 
\ref{fig1a},
we take the initial value of $\ln k_i = -11.8$, which corresponds to times several
e-foldings before horizon crossing, and integrate to a final value of $\ln k_f = -1.4$.
The derived potentials, and the range over which $\phi$ changes, depend strongly
on the choice of $\cH_0$.  However, fairly universal behavior can be seen by graphing
the function
\beq
\widetilde{\cal V}(\phi) = 
 { {\cal V}(\phi/\phi_{\rm f}) - {\cal V}_{\rm f} \over {\cal V}_{\rm i} - 
{\cal V}_{\rm f}}
\eeq
where $\phi_{\rm f}$ is the final value of $\phi$, and ${\cal V}_{\rm i,f}$
are respectively the initial and final values of ${\cal V}$,
which is a monotonically decreasing function.  This is
shown in figure \ref{reconV}.  The rather narrow range of potential shapes shown 
are actually inclusive, since potentials with smaller values of $\cH_0$ are very 
close to the curve for $\cH_0=0.01$, while the largest possible value of $\cH_0$
(given by $\sqrt {{\cal P}{(k_i)}}\cong 0.215$ using the spectrum (\ref{power}))
yields a potential nearly coinciding with that of $\cH_0=0.2$.  Notice that
$\cH_0$ is not allowed to exceed ${\cal P}(k_i)$ in (\ref{scaled}).  This traces
back to the physical restriction that the Hubble parameter can never increase during
inflation, or equivalently, the kinetic energy of the scalar field cannot be negative.

\DOUBLEFIGURE[h]{upot4.eps,width=1.0\hsize}{V-re2.eps,width=\hsize}{Reconstructed potentials for a range
of initial Hubble parameters.  ``Corrected'' curves are discussed in section 3.3.
\label{reconV}}{Reconstructed 
potentials of different initial Hubble parameters, integrating
over 20 e-foldings of inflation.\label{V-re}}


A much greater variation in the shapes of the rescaled potential can be observed if we
insist on maintaining the running index power spectrum over a longer period of inflation,
going beyond that which affects the observable multipoles of the CMB temperature
fluctuations.  
Figure \ref{V-re} shows that the relative shapes
of inflaton potential in different initial conditions ($\ln k_i= -11.8$,
$\ln k_f = 8.8$). We can see that they are quite different, but
all of them have a ``bump." (Potentials with ${\cal H}_0$ smaller than 0.01 are close to the curve
of ${\cal H}_0=0.01$.)

\EPSFIGURE[h]{params2.eps,width=3.75in}{Dependence of potential parameters
$\phi_{\rm f}$, ${\cal V}_{\rm i}$, and $({\cal V}_{\rm i} - 
{\cal V}_{\rm f})$, 
on the initial Hubble constant.  Equations are numerical fits to the small-$\cH_0$
parts of the curves.
\label{params}}

Although the relative shape of the inflaton potential is somewhat insensitive to
$\cH_0$ (at least over shorter periods of inflation), the scales over which it changes do depend strongly on $\cH_0$.  The
dependences 
\beq
\phi_{\rm f}\sim \cH_0; \quad {\cal V}_{\rm i}\sim \cH_0^2; \quad
{\cal V}_{\rm i} - 
{\cal V}_{\rm f} \sim \cH_0^4
\eeq
are shown in figure \ref{params}.  These dependences can be understood from equations
(\ref{HJ}-\ref{Pk0}).  Since $P(k)$ does not vary, eq.\ (\ref{Pk0}) implies the
scaling $H' \sim H^2$.  Thus eq.\ (\ref{phid}) predicts that $\phi$ changes by 
an amount $\Delta\phi \sim H^2 \Delta t \sim H$, since we are interested in time
scales of order $1/H$.  Eq.\ (\ref{HJ}) obviously predicts that the magnitude 
of $V$ scales like $H^2$.  The variation in $V$ is $\Delta V \sim V' \Delta\phi
\sim H H' \Delta\phi \sim H^4$. 

The numerical value of $\cH_0$ has no physical significance except in
comparison to the value of $\sqrt{\cP}$ at the chosen starting time ({\it i.e.}, starting value of
$\ln k$).  Suppose we evolve the system for two different initial conditions,
$(\ln k_1,\ \cH_1)$ and $(\ln k_2,\ \cH_2)$ with $k_1 < k_2$.  Given the solution for
$(\ln k_1,\ \cH_1)$, it is always possible to find $(\ln k_2,\ \cH_2)$ such that the two
solutions match in the region where they overlap.  However the converse is not true.
This is because of the attractor nature of inflationary solutions: since the effects of
variation of initial conditions are exponentially suppressed with time, it is not generally
possible to find initial conditions at very early times which match an arbitrary deviation
from slow-roll behavior at later times.  This point is important if inflation did not 
start much earlier than horizon crossing, so that the transient effects of initial conditions
can be visible in the low-$k$ part of the power spectrum, as we will discuss in section
\ref{sic}.

\subsection{Slow-Roll Corrections}
\label{src}
In the previous subsections we reconstructed the potential using the
leading slow-roll approximation for the power spectrum.   Here we
show how corrections which are subleading in the slow-roll expansion
can be incorporated in a simple way.\footnote{
Recently ref.\ \cite{Makarov} investigated the accuracy of the
slow-roll expansion for quantities like $n_s$ and $d n_s/d\ln k$
in models with large running, finding that the inclusion of higher
order corrections did not improve the fit of the approximate spectrum
to the exact one.  However, this leaves open the question as to
whether the failure is with the approximation (\ref{stewart}), or 
with its Taylor expansion in powers of $\ln k$.}
The next order result in the slow-roll approximation is
\cite{Stewart}
\beq
\label{stewart}
	{\cal P}^{(2)} (k) = 
\left[ 1- \epsilon + (2- \ln 2 -\gamma) 
(2 \epsilon - \eta) \right] ^2 {\cal P}^{(0)} \equiv {\cal C}\,{\cal P}^{(0)}
\eeq
where $\gamma \simeq 0.5772$ is the Euler-Mascheroni constant, ${\cal P}^{(0)}$
is the lowest order power spectrum which we already computed.
To incorporate these corrections, which are dominated by $\eta$, we make the replacement
${\cal P}\to {{\cal P} / {\cal C}}$ in eqs.\ (\ref{scaled}), where the correction factor ${\cal
C}$ is computed using the previously obtained solution for ${\cal H}$.  This procedure can
be iterated until the solution converges.  Convergence to a self-consistent solution 
is obtained after three iterations. In this way we avoid having to solve a higher order
differential equation for ${\cal H}$.  The second derivatives $\cHodd$, $\podd$ in $\eta$
can be computed in terms of $\cHod$, $\pod$ using the equations of motion (\ref{scaled}).
The result is that the shapes (and magnitudes) of the potentials are slightly modified, as
shown by the ``corrected'' curves in figure\ \ref{reconV}.  The maximum value of $\cH_0$ is
also reduced from $0.21$ to $0.11$.  The changes to the parameters in figure
\ref{params} are so slight that we do not show them.  

\subsection{Sensitivity to Initial Conditions}
\label{sic}

Once the potential has been determined, the power spectrum does not uniquely follow.
One must also supply the initial velocity of the inflaton field.  Normally this is
unimportant since the inflaton reaches the attractor solution in which the slow-roll
approximation $3 H\dot\phi = -V'$ is satisfied, hence $\dot\phi$ is determined by the 
potential.  However the initial velocity required to match the desired power spectrum
does not generally coincide with $-V'/3H$; instead it is given by eq.\ (\ref{Pk0}).

The effect is most pronounced for large values of $\cH_0$, as
illustrated in figure \ref{reconP}.  In this example, (where $\ln k_i
= -9$ and $\cH_0 = 0.55$, and we do not include slow-roll
corrections) the initial value of $\dot\phi$ implied by eq.\
(\ref{Pk0}) is $\dot\phi = 0.76$ (in Planck units), while the
slow-roll value is $\dot\phi = 0.33$.  Using the latter value, the
spectrum is distorted in its shape at the beginning of inflation, and
by a constant horizontal shift $\Delta\ln k$ thereafter. The figure
shows that compensating for $\Delta\ln k$ causes the two spectra to
match well once the transient effect of the initial condition has
died away.  For smaller inflationary scales (small $\cH_0$), the
shift $\Delta\ln k$ becomes negligibly small, while the low-$k$
distortion remains significant.  Since the horizon-crossing condition
is that $\ln k = \ln a + \ln H$, this shift in $\ln k$ can be absorbed
in a change of the scale of inflation.

\EPSFIGURE[h]{ic1.eps,width=0.7\hsize}{Power spectra obtained from reconstructed
potential, for different initial conditions of $\dot\phi$.
\label{reconP}}

\section{Tensor Contribution to CMB}
\label{tc}
It was noted in section \ref{rps} that the power spectrum with large running
provides a good fit to the data only if a large tensor component is also allowed. 
A degeneracy between these parameters is 
understandable, since large running suppresses power at low $k$, while the tensor
contribution enhances it.  In
single-field inflation, which we are assuming, 
the tensor power is related to the scalar power, since at lowest order
in the slow-roll approximation, 
\beq
\label{Pt0}
	P_t(k) = {8 \over c_1} H^2
\eeq
(compare to the scalar power, eq.\ (\ref{Pk0})).  This convention for
the normalization of $P_t$ is consistent with the definition of 
the tensor-to-scalar ratio $r$ used by CAMB and by ref.\ \cite{Peiris},
\beq
	r = {P_t(k_0)\over P(k_0)} = 16\epsilon(k_0).
\label{ratio}
\eeq

Parameterizing the scalar spectrum as eq.\ (\ref{power}) and using the WMAP parameters in
table \ref{tab3}, we can determine the tensor spectrum as well as the reconstructed
potential uniquely.  However, a problem arises with the running spectrum when we try to
achieve a large tensor ratio $r\cong 0.5$ at the given pivot point $k_0$.   Namely, eq.\
(\ref{ratio}) shows that $\epsilon \sim 1/32$ at $k_0$; however if we try to evolve
backwards in time from this point using the running spectrum,  the slow-roll parameters
diverge, and prevent us from extending the initial value of $\ln k$ to as small values as
desired.  This behavior is illustrated in figure \ref{eps1}, which shows the reconstructed
slow-roll parameters corresponding to the WMAP running spectrum.  The figure shows that 
$\epsilon$ remains small ($< 0.05$) from $\ln k \simeq -7.2$ to $1.3$, $\eta$ runs linearly
and gives the running of the power spectrum. But demanding a large tensor-to-scalar requires
the slow-roll parameters to be quite large at the lower scales near $k \sim 10^{-4}{\rm
Mpc}^{-1}$. Moreover, negative running at large $k$ makes the power spectrum decrease
exponentially, again leading to an increase in the slow-roll parameters,  since $\epsilon =
{\cal H}^2/{\cal P}$.  The slow-roll approximation breaks down  before the end of inflation.
For example, we get only 15 e-foldings while $\epsilon<0.2$ in the WMAP running model.
These observations are consistent with ref.\ \cite{CR}, who noted that $\epsilon$ reaches
a minimum near the scale where $n_s$ crosses 1 in models with large running.

To alleviate the above problem, and extend the duration of inflation before the
pivot point, we use the partial running spectrum with a low-$k$ cutoff
which was discussed in section
\ref{rps}, and shown (together with the constructed tensor spectrum) in figure
\ref{figPsPt}. We could also include the high-$k$ cutoff on the running, but we
choose to omit it, in order to discuss the constraints which arise
 on the length of inflation 
if the spectrum continues to run at high $k$.
The tensor spectrum is not a constant as is usually assumed when computing
the CMB temperature anisotropy, and we have modified CosmoMC to use the constructed
tensor spectrum.  However, we do not observe any effect on the likelihood of the best
fit running spectrum, nor any change in the other cosmological parameters,
relative to the standard form of the tensor spectrum.

\DOUBLEFIGURE[h]{eps3.eps,width=1.0\hsize}{lnPt.eps,width=\hsize}{The reconstructed 
slow-roll parameters for WMAP running power spectrum, showing singular behavior at
low $k$.
\label{eps1}}{The partial running
 scalar spectrum and the reconstructed tensor spectrum (unnormalized, $A=1$).\label{figPsPt}}

With the low-$k$ cutoff on the running of the spectrum, the beginning of inflation can
be pushed back before horizon crossing, while maintaining a large tensor ratio at the
pivot point; this is shown in figure \ref{eps2}, where the slow-roll parameters are seen
to remain small during the relevant part of inflation, while the minimum value of $\epsilon$
is consistent with $r\cong 0.5$.
The discontinuity in $\eta$ is due to the discontinuity in the slope
of the power spectrum where it goes between nonrunning and running, 
which could be avoided by making a smoother transition.
On the other hand, since we chose not to cut off the
running at high $k$, there is a limit on the amount of inflation after horizon
crossing.  The reconstructed potential continues to decrease monotonically, so there
is no abrupt end to inflation, but $\epsilon$ becomes steadily larger, and quasi-exponential
inflation crosses over to power-law inflation with a small power.
This model gives 16 e-foldings of exponential inflation, during which
$\epsilon < \epsilon_i = 0.12$.  The transition to reduced acceleration is shown in 
figure \ref{lna}, where one can see a break in the logarithm of the scale factor
versus time at around $\ln a = 20$ e-foldings. {
Figure \ref{lna3}, which plots $\ln a$ versus $\ln t$, shows that
after the break the expansion is indeed power-law.}

\EPSFIGURE[h]{eps5.eps,width=3.75in}{The reconstructed 
slow-roll parameters for partial running power spectrum.\label{eps2}}

\DOUBLEFIGURE[h]{lna.eps,width=\hsize}{lna3.eps,width=1.0\hsize}{The logarithm 
of the scale factor as a function of time, showing the transition from exponential
inflation to power-law expansion.\label{lna}}{Same as figure 12, but
plotting $\ln a$ versus $\ln t$, to show the 
power-law character of the expansion at late times.
\label{lna3}}

With the given tensor ratio, we can uniquely determine the scale of inflation and the
single out one reconstructed potential from the family that was derived in section
\ref{RP}.  Members of this family were parametrized by the ratio
\beq
	{{\cal H}_0^2\over {\cal P}} =
	{{H^2}\over 2c_1\,P}
	= \frac{1}{16} {P_t\over P} = {r\over 16} = {\epsilon} 
\eeq
evaluated at the {\it initial} value of $\ln k$ (not the pivot point).  Figure \ref{eps2} shows that $\epsilon\cong 0.12$ at the initial
value $\ln k_i\cong -11.8$, in order to get a large enough tensor ratio at the pivot point.
We find that ${\cal H}_0\cong 0.29$, which {satisfies} 
the constraint ${\cal H}^2\le \cP$
in the model with the partial running spectrum.
We can then determine the scale of inflation;
from {table \ref{tab2}}, $A=2.1 \times 10^{-9}$, we obtain 
 $\alpha = \sqrt{2 c_1 A} \simeq 4.1 \times 10^{-4} M_{\rm Pl} \simeq 9.9
\times 10^{14}{\rm GeV}$ (see eq.\ (\ref{alphaeq})).  This corresponds to a Hubble scale
of $H=\alpha\cH_0 = 2.9\times 10^{14}$ GeV at the beginning of inflation, and 
an energy scale of $V^{1/4} = (3H^2M_{\rm Pl}^2)^{1/4} = 3.5\times 10^{16}$  GeV.
This is the energy scale expected in the slow-roll
approximation with $\epsilon \simeq 0.1$, $V^{1/4} \simeq 0.027 \epsilon^{1/4} M_{\rm Pl} \simeq 3.7 \times 10^{16}$GeV.
The reconstructed Hubble parameter and potential, as functions of $\phi$,
are shown in figures
\ref{Hphi} and \ref{Vphi}, where we contrast the predictions of the leading order slow
roll approximation and those including corrections at second order in the slow-roll
parameters.  The corrections are seen to be a small effect.

\DOUBLEFIGURE[h]{Hphi3.eps,width=1.0\hsize}{Vphi3.eps,width=\hsize}{The reconstructed 
Hubble parameter, at first and second order in the slow-roll approximation, 
for the partial running spectrum.\label{Hphi}}{The reconstructed potential, 
at first and second order in the slow-roll approximation, for the partial running spectrum.\label{Vphi}}

The reconstructed potential shows the ``bump'' which is characteristic of models with
large running {from blue ($n>1$) to red ($n<1$)}, since a large third derivative of the potential is required to get large
running in the slow-roll prediction, 
\beq
	{dn_s\over d\ln k} = -2\xi_{\rm V} + 16\epsilon_{\rm V} \eta_{\rm V}  - 24\epsilon_{\rm V}^2
\eeq
in terms of the usual potential slow-roll parameters: $\epsilon_{\rm V} = \frac12 (V^{\prime}/V)^2$, $\eta_{\rm V} = V^{\prime \prime}/V$, and $\xi_{\rm 
V} = V'V'''/V^2$.  From the perspective
of particle physics model-building, it is interesting to ask to what extent this potential
can be matched by a renormalizable potential.  We have therefore fit the reconstructed
$V$ to a quartic polynomial.  Obviously the fit becomes better as one restricts the
range of the field, so we have done this exercise for a series of final $\phi$ values, which
correspond to different maximum wave numbers in the spectrum.  We made our fits using
$\ln k_i= -9.2$ ($10^{-4}$ Mpc$^{-1}$) to $\ln k_f = 0$ (1 Mpc$^{-1}$), 3.8, 4.8, and 5.8,
obtaining fits which for the case of $\ln k_f=4.8$ has the form 
\beq
V = 1.7 \times 10^{-7} \times (0.277 - 0.163\, \phi + 0.0583\, 
\phi^2 - 0.0102\, \phi^3 + 0.00064\,  \phi^4)
\eeq
where all coefficients (and $\phi$) are in Planck units.
Figure \ref{Vfit} plots the approximate fits against the actual reconstructed potential.
Plotted in this way, all the fits look very good, but
  to better discriminate, we show the
fractional deviations $\Delta V/V$ {and correlation coefficients} between the fits and the actual potential in figure
\ref{Verror}.  We see that final $k$ values up to $\ln k_f=4.8$ provide good fits at the
percent level.  This $\ln k_f$ value corresponds to the first 18 e-foldings of inflation.
The difficulty of sustaining $50-60$ e-foldings of inflation with large running has
been noted previously \cite{CST}.  Our analysis gives further 
support to the idea that large running
requires some significant change in the potential after the first few e-foldings, such as
can occur in multifield models \cite{easther, yam-yok}.  It is also noteworthy 
in our reconstruction that the inflaton changes by superPlanckian values during this
initial period of inflation.  However, the smallness of the coefficients in the potential
indicate that the effective field theory description is not invalidated by the large
field values.

\DOUBLEFIGURE[h]{Vfit.eps,width=1.0\hsize}{Verror.eps,width=\hsize}{The
reconstructed inflaton potential, and a series of renormalizable 
approximate potentials, which are matched to the exact result over
increasingly large ranges of field values (indicated by the wave
number of horizon crossing corresponding to the maximum field
value in each range).\label{Vfit}}{
The fractional errors of the fitting potentials.  The correlation
coefficient $r$ for each fit is also shown next to the label ($r$
close to 1 corresponds to a good fit).
\label{Verror}}

\section{Conclusions} \label{conc} We have revisited the question of a running spectral
index, motivated by the WMAP first year data.  We are allowing for the possibility that the
large running  applies only in a limited region of $k$ space, where the CMB experiments are
most  sensitive and large scale structure is less sensitive.  It is possible that even a
parametrization with a running spectral index does not provide a good representation of the
actual power spectrum over all observable scales.  In fact we know that inflation cannot
last for 60 e-foldings if running remains at the level of $dn/d\ln k = -0.1$---we find only
$15-20$ e-foldings of exponential inflation---so it is also theoretically well-motivated to
fit the spectrum separately in low and high regions of $k$ space.  Something must change
at larger $k$ just so that inflation can continue for another $40-45$ e-foldings.

Although the evidence for large running from the CMB is marginal, an
improvement in $\chi^2$ of 3, requiring the addition of two (although
somewhat degenerate) parameters, we have argued that it is  based on
more than just the first few multipoles of the temperature
anisotropy; in fact, including the TE polarization data makes the
first few multipoles unfavorable toward running.  Rather, one
observes a steady accumulation of $\Delta \chi^2$ as more {low}
multipoles are {dropped}, with a sharp drop occuring in the region
of  $l$ near $40$.  Even though this is still in the region where
errors in the WMAP data  are limited by cosmic variance, it will be
interesting to see if there any significant changes when the next two
years of data will be released.\footnote{We are posting this paper on
the astro-ph archive a day ahead of the second WMAP release,
announced for 16 March 2006.}

Even though the large running scenario may be unlikely, it is interesting to pursue its
consequences, given that it is not ruled out by the data.  Since the slow-roll parameters do
not remain very small over the entire inflationary trajectory, it may be useful to have a
method of reconstruction of the inflaton potential which is not based on the slow-roll
expansion around a particular pivot point, but instead satisfies the slow-roll prediction
for the power spectrum (computed to any desired order in the slow-roll expansion) at {\it
every} point in $\phi$ or $k$ space.  We have devised 
{an integration} method
of reconstructing $V(\phi)$ directly from $P(k)$, based on the Hamilton-Jacobi formalism,
which has this property, and is also simple to implement numerically.  We have shown how
corrections to any order in the slow-roll expansion can be implemented without significantly
complicating the algorithm.  This method may be of interest in itself, independently of 
whether the running spectrum is confirmed by future data.

We demonstrated the reconstruction method by deriving the family of inflaton potentials that
reproduce the best-fit large running spectrum.  We subsequently focused on the unique
potential which gives a large enough tensor-to-scalar ratio $r$ to be consistent with the
best-fit spectrum.  We pointed out that this is only possible when the running is cut off at
low $k$ values; otherwise $r$ itself runs toward smaller values so quickly  with $k$ that it
falls below the desired value $r\cong 0.5$ before the pivot point $k_0$ is reached.  This is
a new observation, though anticipated by \cite{CR}.  The large tensor contribution is one of
the interesting features of the large running model; it puts the scale of inflation as high
as possible (we find $V^{1/4} =  3.5\times 10^{16}$  GeV) and it provides the  greatest hope
that direct evidence of tensor modes might be seen in the CMB.  It also provides the best
chance of uniquely fixing the inflaton potential, since without the tensor spectrum, and in
the absence of a specific model, it is always possible to lower the scale of inflation while
simultaneously making it flatter, to keep the scalar power fixed. We have shown that a
renormalizable potential provides a good fit to the actual reconstructed potential over the
limted region of field space where inflation occurs, even though the field values 
are Planckian.  This provides another intriguing aspect to the large-running model:
if true, it may give us an inflationary window on physics at the Planck scale.

{\bf Acknowledgments.}  We thank Gil Holder, Antony Lewis, Andrew Liddle, and Licia Verde
for helpful information, discussions, or advice about CosmoMC.   L.~Hoi is supported by a 
postgraduate scholarship of the Tertiary Education Services Office, Macao SAR.
We are also supported by NSERC of Canada and FQRNT of Qu\'ebec.

\end{document}